\documentclass[twocolumn,showpacs,preprintnumbers,amssymb]{revtex4}

\usepackage{graphicx,epsf, epsfig, psfig, amssymb}
\usepackage{bm}

\def\be{\begin{equation}}
\def\ee{\end{equation}}
\def\beq{\begin{eqnarray}}
\def\eeq{\end{eqnarray}}

\def\IL{\relax{\rm I\kern-.18em L}}

\begin{document}

\title{What we (don't) know about black hole formation in high-energy
collisions}
\author{Vitor Cardoso}
\email{vcardoso@wugrav.wustl.edu} \affiliation{McDonnell Center for
the Space Sciences, Department of Physics, Washington University,
St.\ Louis, Missouri 63130, USA \footnote{Also at Centro de
F\'{\i}sica Computacional, Universidade de Coimbra, P-3004-516
Coimbra, Portugal}}

\author{Emanuele Berti}
\email{berti@wugrav.wustl.edu} \affiliation{McDonnell Center for the
Space Sciences, Department of Physics, Washington University, St.\
Louis, Missouri 63130, USA}

\author{Marco Cavagli\`a}
\email{cavaglia@olemiss.edu} \affiliation{Department of Physics and Astronomy,
University of Mississippi, University, MS 38677-1848, USA}

\date{\today}

\begin{abstract}
Higher-dimensional scenarios allow for the formation of mini-black holes from
TeV-scale particle collisions. The purpose of this paper is to review and
compare different methods for the estimate of the total gravitational energy
emitted in this process. To date, black hole formation has mainly been studied
using an apparent horizon search technique. This approach yields only an upper
bound on the gravitational energy emitted during black hole formation.
Alternative calculations based on instantaneous collisions of point particles
and black hole perturbation theory suggest that the emitted gravitational
energy may be smaller. New and more refined methods may be necessary to
accurately describe black hole formation in high-energy particle collisions.
\end{abstract}


\maketitle
\section{Introduction}
The standard model of particle physics has been successfully tested
up to energies of $\sim 1$ TeV. However, its foundations are still
mysterious. Much effort has been spent to explain the hierarchy
problem, i.e.\ the huge difference between the electroweak scale,
$m_{\rm EW} \sim 300\, {\rm GeV}$, and the Planck scale, $M_{\rm Pl}
\sim 10^{19} \,{\rm GeV}$. While electroweak interactions have been
probed at distances $m_{\rm EW}^{-1} \sim 10^{-16}\, {\rm cm}$,
gravitational forces have not been probed at distances $M_{\rm
Pl}^{-1}\sim 10^{-33} \, {\rm cm}$. Gravity has only been accurately
measured in the $\sim 0.01$ cm range \cite{Hoyle:2004cw}.

If gravity is modified at scales smaller than $1 \,{\rm mm}$, the hierarchy
problem can be solved by assuming the existence of $n$ compact extra dimensions
of length $\sim R$ \cite{hamed}. Gauss's law in $D=4+n$ dimensions implies that
two test masses $m_1$, $m_2$ at a distance $r \ll R$ feel a gravitational
potential
\begin{equation}
V(r)=G_{D}\frac{m_1 m_2}{r^n r}\,\,,\,\,\,\,r \ll R\,,
\label{gaussndim}
\end{equation}
where $G_{D}$ is the $D$-dimensional Newton constant. The four-dimensional
Newtonian potential is recovered at distances $r\gg R$, where $r^n\sim R^n$ and
$G_4=G_{D}/R^n$. The $D$-dimensional Planck mass, $M_{\rm Pl,D}$, is obtained
by equating the Schwarzschild radius of an object of mass $m$ to its Compton
wavelength, $\lambda=1/m$:
\begin{equation}
(M_{\rm Pl,D})^{D-2}\sim\frac{1}{G_{D}}\,. \label{compt1}
\end{equation}
(Here and throughout the paper we set $\hbar=1$ and $c=1$.) $M_{\rm Pl,D}$ is
related to the effective four-dimensional Planck scale $M_{\rm Pl}$ by $M_{\rm
Pl}^2 \sim (M_{\rm Pl,D})^{D-2}R^n$. The hierarchy problem is solved by
imposing equal scales in the higher-dimensional setting, i.e., $ m_{\rm EW}=
M_{\rm Pl,D}$. This condition relates the size $R$ of the extra dimensions to
the number of extra dimensions $n$. A single extra dimension implies deviations
from Newtonian gravity over solar system distances. It is thus excluded
empirically. If $n=2$, the size of the extra dimensions is $R\sim 0.3\,{\rm
mm}$. This value has been recently ruled out by experiments with torsion
pendulums \cite{Hoyle:2004cw}. However, any $n \geq 3$ gives modifications of
Newtonian gravity at distances smaller than those currently probed by
experiment.

An exciting consequence of TeV-scale gravity is the possibility of production
of black holes (BHs) in particle colliders \cite{bhprod,bhprodreviews} and
ultra high energy cosmic ray interactions with the atmosphere
\cite{blackholescosmicrays}. A naive estimate of the cross section for $M_{\rm
Pl,D} \sim 1\,{\rm TeV}$ predicts that super-TeV particle colliders will
produce BHs at a rate of few per second.

The BH starts to decay after forming. First, it radiates all the
excess multipole moments. The BH has initially a highly asymmetric
shape. It then settles down to a stationary state. (In four
dimensions it must be a Kerr-Newman BH, by the uniqueness theorem.)
This phase is commonly called {\it balding phase}. The endpoint of
the balding phase is a spinning BH. This BH starts to lose angular
momentum ({\it spin-down phase}). Page \cite{page} has shown that a
spinning BH radiates mainly along the equatorial plane, and that
this radiation carries away most of the BH angular momentum. After
completing the spin-down phase, the BH radiates its remaining
degrees of freedom through Hawking radiation ({\it Schwarzschild
phase}) \cite{pagehigherd,park}. The endpoint of the Schwarzschild
phase is either explosive or a BH remnant is left \cite{remnant}.

The focus of this paper is on the formation of a BH in high-energy
particle collisions. In particular, we are interested in the total
gravitational energy radiated. Different methods have been discussed
in the literature. The standard procedure is to describe the
incoming particles by two Aichelburg-Sexl shock waves
\cite{aichelburg}, and then find a closed trapped surface in the
union of these shock waves
\cite{D'Eath,payne,eardley,yoshino,yoshinorychkov,giddingsrychkov,trapsurfaces,rychkov}.
The shock wave approach can be used to find lower bounds on the mass of
the newly formed BH, i.e., upper bounds on the amount of gravitational
radiation emitted during the process. However, there is little
discussion in the literature on how well these results approximate the
actual gravitational emission. The aim of this paper is to provide a
critical discussion of the different methods. In Section \ref{as} we
describe briefly the standard Aichelburg-Sexl shock wave technique,
the main results of this approach, and its shortcomings. In Section
\ref{alternative} we discuss two alternative techniques. Conclusions
are presented in Section \ref{concl}.
\section{\label{as} Black hole formation via Aichelburg-Sexl shock waves}
Consider a $D$-dimensional Schwarzschild-Tangherlini solution
with mass $M$ \cite{tangherlini}. The metric in spherical coordinates is
\beq ds^2& &= -\left(1-\frac{16\pi G_{D}
M}{(D-2)\Omega_{D-2}}\frac{1}{r^{D-3}}\right)dt^2+\nonumber \\ & &
\left(1-\frac{16\pi G_{D}
M}{(D-2)\Omega_{D-2}}\frac{1}{r^{D-3}}\right)^{-1}dr^2 +r^2
d\Omega_{D-2}^2\,, \label{metricmyersperry} \eeq
where $\Omega_{D-2}$ is the volume of the unit $(D-2)$-dimensional sphere
\begin{equation}
\Omega_{D-2}=\frac{2\pi^{(D-1)/2}}{\Gamma[(D-1)/2]}\,,
\label{integratedsolidangle}
\end{equation}
and $\Gamma[z]$ is Euler's Gamma function. Boosting this solution to
very large values of the Lorentz factor $\gamma$ at fixed total energy
$\mu=M\gamma$, we obtain the Aichelburg-Sexl solution
\be ds^2=-d
\bar{u}d\bar{v}+(d\bar{x}^i)^2+\Phi(\bar{x}^i)\delta (\bar{u})d\bar{u}^2 \,,
\label{aichsexl}\ee
where $\Phi$ depends only on the transverse radius
$\bar{\rho}^2=\bar{x}^i\bar{x}_i$:
\beq \Phi&=&-8G_{D}\mu \log{\bar{\rho}}\,\,,\,\,D=4 \nonumber \\
\Phi&=&\frac{16\pi G_{D}\mu}{\Omega
_{D-3}(D-4)\bar{\rho}^{D-4}}\,\,,\,\,D>4\,.\eeq
Equation (\ref{aichsexl}) describes a particle with energy $\mu$ moving in the
$+z$-direction. The spacetime is flat outside the null plane $\bar{u}=0$. If we
consider an identical shock wave traveling along $\bar{v}=0$ in the $-z$
direction, the two solutions can be superposed to yield a solution for two
colliding shock waves. BH formation can be studied by identifying a future
trapped surface in the solution, with no need to calculate the gravitational
field
\cite{D'Eath,payne,eardley,yoshino,yoshinorychkov,giddingsrychkov,trapsurfaces,rychkov,veneziano}.
If the collision is head-on, it is easy to find a trapped surface in the union
of two flat disks with radii
\be \rho _c=\left (\frac{8\pi G_{D}\mu}{\Omega _{D-3}}\right
)^{1/(D-3)}\,. \ee
Since the BH horizon is always in the exterior region, the method gives a lower
bound on the final BH mass. If the final BH is non-spinning, its mass is
\be
M_{BH}\gtrsim  \mu\left (\frac{(D-2)\Omega
_{D-2}}{2\Omega _{D-3}}\right )^2 \,.\label{masslower}\ee
The total gravitational energy radiated is $2\mu-M_{BH}$, and the
efficiency is $\epsilon \equiv 1-M_{BH}/2\mu$. An efficiency
of $0$\% implies that no gravitational radiation is emitted. An
upper limit on the efficiency follows from Eq. (\ref{masslower}).
Some values are shown in Table \ref{tab:efficiency1}.

\begin{table}
\caption{\label{tab:efficiency1} Upper limits on the efficiency of
gravitational radiation emission for different spacetime
dimensions $D$, using the trapped surface method.}
\begin{ruledtabular}
\begin{tabular}{cc}  \hline
$D$ & $\epsilon$ (\%)\\
\hline 4   &   29.3  \\ \hline 6   &  36.1
\\ \hline 8   &  39.3    \\ \hline 10  & 41.2 \\ \hline $\infty$  & 50  \\
\hline
\end{tabular}
\end{ruledtabular}
\end{table}

The above formalism can be extended to the study of collisions with
nonzero impact parameter
\cite{eardley,yoshino,yoshinorychkov,trapsurfaces}. This makes
possible the computation of the BH production cross-section.
\subsection{\label{upperlimits} Accuracy of the results}
The trapped surface method gives only a lower bound on the BH mass. The BH mass
can be anything between this bound and the center of mass (c.m.) energy of the
collision. D'Eath and Payne studied this problem in more detail for the
four-dimensional case \cite{payne}. Bondi's news function, which describes the
emission of gravitational radiation, can be written as an infinite series
around the collision axis. The first term of the series is
Eq.~(\ref{masslower}). Making some assumptions on the angular dependence of the
radiation, and extrapolating off the axis, the second term is found to decrease
the efficiency for gravitational wave generation in head-on collisions to
$16$\%. Although this derivation relies on various approximations, the
reduction in efficiency is significant, and may signal that the series is
slowly converging. The situation seems to get even worse in higher dimensions
(see Section \ref{alternative}) and for large impact parameters. As an
illustration of how the trapped surface method may lead to an inaccurate
estimate of the BH mass, let us consider the collision of two non-spinning BHs
initially at rest. Using similar arguments to the trapped surface method,
Hawking \cite{hawking} placed an upper limit of $29.3$\% on the efficiency of
gravitational wave generation. The exact result can be obtained through a
numerical solution of Einstein's equations
\cite{smarr1,smarr2,anninos,anninos-pert,boosted,boosted2,sperhake}, and is
around $0.1$\%, i.e.\ two orders of magnitude smaller than the Hawking bound.
The results for ultra-relativistic collisions are likely to be more accurate.
However, the total energy radiated may still be much smaller than the upper
limits of Table \ref{tab:efficiency1}.
\subsection{\label{pointparticle} Finite size particles}
The Aichelburg-Sexl solution describes the metric of a massless pointlike
particle with a very large boost. Classically, the point-particle assumption is
accurate along the collision axis because of the large Lorentz contraction due
to the boost. However, it fails for directions transversal to the motion. If
the colliding particles are strings, string size effects can be modeled by
considering beam-beam collisions, with beam sizes of order $\lambda _S$, where
$\lambda _S\gtrsim l_{Pl}$ \cite{veneziano}. The effects of finite-size
transversal dimensions have been studied analytically by Kohlprath and
Veneziano \cite{veneziano}, and lead to a smaller cross section. (We are not
aware of any numerical study.) For a reasonable cross section the c.m.\ energy
should satisfy $E > M_{Pl}(\lambda _S/l_{Pl})^{D-3}$.
\subsection{\label{nonspinning} Spin effects}
In general, colliding particles have intrinsic spin and should be
modeled by metrics other than the Tangherlini metric. A naive
generalization of the Aichelburg-Sexl approach is to boost the
rotating Myers-Perry metric \cite{myersperry}. This has been done by
several authors \cite{boostkerr}. The results are cumbersome enough to
make the trapped surface method very difficult to implement. A
seemingly better candidate to model spinning high-energy particles was
recently proposed in Ref.~\cite{frolov}. This model consists in a
spinning radiation beam-pulse which includes dragging effects, in
contrast to the boosted Kerr metric.

Another question concerns the angular momentum of the BH, given the
intrinsic spin and angular momentum of the colliding particles.
(This is also an important point in astrophysics \cite{merrit}.) The
cross section and the angular momentum of the final BH can be
estimated by assuming the net angular momentum carried by
gravitational waves to be negligible \cite{park,eardley}. However, a
more accurate investigation of angular momentum effects is needed.
\subsection{\label{charge} Charge effects}
String length and spin effects should be important if the incoming
particles have energy close to the Planck scale. It is likely that
both of these effects are suppressed for super-Planckian
energies. Charge effects are expected to dominate at very high
energies because gauge fields are confined on the brane and decay more
slowly than the gravitational field of a neutral particle
\cite{yoshinosuggestion}. Estimates of charge effects in BH
production from high-energy collisions are not yet available in the
literature. A first attempt was presented in Ref.~\cite{cardosocharge}
using a perturbative method.
\section{\label{alternative}Other methods to estimate the energy released}
Because of the problems listed above, it is desirable to explore different
methods. In this section we discuss two possible approaches. Although they are
not free from shortcomings, quantitative results on the BH formation
process can be obtained. These techniques give consistent results in $D=4$.

The formalism to handle gravitational waves in $D$-dimensional flat
spacetimes is discussed in Ref.~\cite{cardosoDwaves}. The nonlinearity
of Einstein's equations makes the treatment of the gravitational
radiation problem difficult.  It is a standard procedure to work only
with the weak radiative solution. In this limit, the energy-momentum
self-interaction term of the gravitational wave can be neglected. This
approach is justified in most problems where the total amount of
gravitational radiation released is negligible in comparison to the
total energy content of the spacetime. (High-energy collisions of two
BHs require in principle the inclusion of nonlinear effects.) Let us
assume an asymptotically flat $D$-dimensional spacetime with metric
$g_{\mu\nu}=\eta_{\mu\nu}+h_{\mu\nu}$. At this linearized level it can
be shown that

{\bf (i)} gravitational waves in a $D$-dimensional spacetime have
$D(D-3)/2$ independent polarizations;

{\bf (ii)} the perturbation amplitude in the wave zone, $h_{\mu \nu}$, can be
expressed in terms of the energy momentum tensor as
\beq h_{\mu\nu}(t,{\bf x})&=& -8 \pi G_{D} \frac{1}{(2\pi
r)^{(D-2)/2}}
\partial_{t}^{(\frac{D-4}{2})}\times \nonumber \\
& & \left[ \int d^{D-1}{\bf x'}
 S_{\mu \nu}(t-|{\bf x-x'}|,{\bf x'})\right]\,,
\label{retsolwavezone} \eeq
where $\partial_{t}^{(\frac{D-4}{2})}$ stands for the
$\frac{D-4}{2}$th derivative with respect to time and
\begin{equation}
S_{\mu\nu}=T_{\mu\nu}-\frac{1}{D-2}\,\eta_{\mu\nu}\,T^{\alpha}_{\;\;\;\alpha}\,.
 \label{S}
\end{equation}
The Fourier transform and the energy spectrum are (see
Ref.~\cite{cardosoDwaves} for details)
\begin{equation}
h_{\mu\nu}(\omega,{\bf x})= -\frac{8 \pi
G_{D}\omega^{(D-4)/2}}{(2\pi r)^{(D-2)/2}}e^{i\omega r} \int
d^{D-1}{\bf x'} S_{\mu \nu}(\omega,{\bf x'})\,,
\label{retsolwavezonefourier}
\end{equation}
\beq & & \frac{d^2E}{d\omega d\Omega}= 2 G_{D}
\frac{\omega^{D-2}}{(2\pi)^{D-4}} \times \nonumber \\
& & \left( T^{\mu\nu}(\omega,{\bf k})T_{\mu\nu}^*(\omega,{\bf
k})-\frac{1}{D-2} |T^{\lambda}_{\:\:\:\lambda}(\omega, {\bf
k})|^2\right)\,, \label{powerwavezone} \eeq
respectively.
\subsection{Instantaneous Collisions in Even $\bm D$-Dimensions}
In general, two scattering bodies release gravitational energy due to
momentum exchange. If the collision is hard, i.e.\ the incoming and
outgoing trajectories have constant velocities, the metric
perturbation and the released energy can be computed exactly. (This
method was first derived by Weinberg \cite{weinberg,wein1} and later
explored in Ref.~\cite{wein2}.) These calculations assume an
instantaneous collision, and are valid for arbitrary velocities and
low energies. The resulting spectrum is flat in four dimensions
\cite{jackson}, and a cutoff frequency is needed to obtain a finite
total energy. A suitable cutoff enables us to estimate the total
energy radiated by the collision of two ultra-relativistic BHs. This
method has been recently generalized to higher dimensions in
Ref.~\cite{cardosoDwaves}. In what follows we revisit this result and
estimate the total energy radiated in the high-energy collision of two
$D$-dimensional BHs.

Consider a system of freely moving particles with $D$-momenta $P_{i}^{\mu}$,
energies $E_i$ and ($D-1$)-velocities ${\bf v}$. These quantities change
abruptly at $t=0$ to corresponding primed quantities due to the collision.
The energy-momentum tensor is
\beq
T^{\mu\nu}(t, {\bf v})&=& \sum_i
\biggl[
\frac{P_{i}^{\mu}P_{i}^{\nu}}{E_i}
\delta^{D-1}({\bf x}-{\bf v}t)\Theta(-t)+\nonumber \\
& & \frac{{P'}_{i}^{\mu}{P'}_{i}^{\nu}}{E'_i} \delta^{D-1}({\bf
x'}-{\bf v'}t)\Theta(t)\,\biggr].
\label{enmomtenpointpctles}
\eeq
Substituting Eq.~(\ref{enmomtenpointpctles}) in
Eqs.~(\ref{retsolwavezonefourier}) and (\ref{powerwavezone}), the
metric perturbation $h_{\mu\nu}$ and the radiation spectrum can be
obtained. Let us consider a head-on collision of a particle with
mass $m_1$ and Lorentz factor $\gamma_1$ with a particle of mass
$m_2$ and  Lorentz factor $\gamma_2$ in the c.m.\ frame. We assume
without loss of generality that the motion is in the $(x_{D-1},x_D)$
plane:
\begin{eqnarray}
P_{1}&=& \gamma_1m_1
(1,0,0,...,v_1\sin\theta_1,v_1\cos\theta_1)\,,\nonumber \\
P'_1&=&(E'_{1},0,0,...,0,0)\,, \label{momenta1}
\\
P_{2}&=& \gamma_2m_2
(1,0,0,...,-v_2\sin\theta_1,-v_2\cos\theta_1)\,,\nonumber \\
P'_2&=&(E'_{2},0,0,...,0,0)\,. \label{momenta2}
\end{eqnarray}
Momentum conservation leads to the additional relation
$\gamma_1m_1v_1=\gamma_2m_2v_2$. Substituting Eqs.~(\ref{momenta1}) and
(\ref{momenta2}) in Eq.~(\ref{enmomtenpointpctles}) and (\ref{powerwavezone}),
we find
\begin{equation}
\frac{d^2E}{d\omega d\Omega}=\frac{2G_{D}}{(2\pi)^{D-2}}
\frac{D-3}{D-2}\frac{\gamma_{1}^2m_{1}^2v_{1}^2(v_1+v_2)^2
\sin^4{\theta_1}\,\omega^{D-4}}{(1-v_1\cos\theta_1)^2(1+v_2\cos\theta_1)^2}\,.
\label{energypersolidanglefreqinstcol}
\end{equation}
For arbitrary (even) dimensions $D>4$ the spectrum is not flat. The integration
of Eq.~(\ref{energypersolidanglefreqinstcol}) from $\omega=0$ to a cutoff
frequency $\omega_c$ gives
\begin{equation}
\frac{dE}{d\Omega}=\frac{2G_{D}}{(2\pi)^{D-2}}
\frac{1}{D-2}\frac{\gamma_{1}^2m_{1}^2v_{1}^2(v_1+v_2)^2
\sin^4{\theta_1}\,\omega_{c}^{D-3}}{(1-v_1\cos\theta_1)^2(1+v_2\cos\theta_1)^2}
\,. \label{energypersolidangleinstcol}
\end{equation}
%
If a BH forms in the collision, the effective timescale for
the process is $\tau \sim r_+$. This suggests to approximate the
cutoff frequency $\omega_c$ by the inverse of the BH radius.
For a collision between equal-mass particles ($m_1=m_2=m$,
$v_1=v_2=v$) this yields a total energy
\be E=\frac{2^{3-D}\gamma ^2 m^2} {M \Gamma ^2[(D-1)/2]} \,,\ee
where $M\sim 2m\gamma$. The efficiency $E/(2m\gamma)$ is
\be \epsilon=\frac{2^{1-D}} { \Gamma ^2[(D-1)/2]} \,.\ee
Values of $\epsilon$ are listed in Table \ref{tab:efficiency2} for
different dimensions. The results are in agreement with the
four-dimensional estimate of D'Eath and Payne \cite{payne}. However,
the total energy decreases with the spacetime dimension, in
disagreement with the estimate of Ref.~\cite{eardley}.

\begin{table}
\caption{\label{tab:efficiency2} Efficiency for gravitational radiation
generation in head-on collisions for different spacetime dimensions $D$, using
the instantaneous collision method.}
\begin{ruledtabular}
\begin{tabular}{cc}  \hline
$D$ & $\epsilon$ (\%)\\
\hline 4   &   15.9  \\ \hline 6   &  1.8
\\ \hline 8   &  0.07    \\ \hline 10  & 0.001 \\ \hline $\infty$  & 0  \\
\hline
\end{tabular}
\end{ruledtabular}
\end{table}

The instantaneous collision approach seems to suggest that the trapped
surface method overestimates the total energy emitted in the
collision. The approach described in this section can be generalized
to describe the collision of rotating bodies. Extension to rotating
BHs would also be of interest for the computation of gravitational
radiation from gamma-ray bursts \cite{piran}.

\subsection{Perturbation around the black hole background}
The collision of a BH with an ultra-relativistic particle was studied in detail
by Cardoso and Lemos \cite{cardosorad} in four dimensions, and by Berti {\it et
al.} for $D>4$ \cite{bertimarcoleonardo}. In the perturbative approach, the
BH-particle system is described by a perturbed Schwarzschild-Tangherlini BH,
where the perturbation $h_{\mu\nu}$ is induced by the infalling particle.
Expanding Einstein's equations to first order in $h_{\mu \nu}$, the problem can
be expressed as a second order differential equation for $h_{\mu \nu}$.
Although the formalism is strictly valid only for a colliding particle with
small energy $E$ compared to the BH mass, the results can be extrapolated to
$E\sim M_{BH}$ \cite{bertimarcoleonardo}.  The results of this analysis are
given in Table \ref{tab:efficiency3}. They are in qualitative agreement with
the predictions of the instantaneous collision approach.
%
\begin{table}
\caption{\label{tab:efficiency3} Efficiency for gravitational radiation
generation in head-on collisions for different spacetime dimensions $D$, using
the perturbed Schwarzschild-Tangherlini BH method.}
\begin{ruledtabular}
\begin{tabular}{cc}  \hline
$D$ & $\epsilon$ (\%)\\
\hline 4   &  13   \\ \hline 6   &10
\\ \hline 8   &   7  \\ \hline 10  &8  \\
\hline
\end{tabular}
\end{ruledtabular}
\end{table}
%

The total energy radiated by a freely-falling particle in the
four-dimensional Schwarzschild spacetime \cite{davis} is in very
good agreement with numerical simulations of BH head-on collisions.
An extensive comparison of numerical and perturbative results can be
found in Refs.~\cite{smarr1,smarr2,anninos,anninos-pert}. According
to perturbation theory, the total energy radiated by a test particle
of mass $m$ falling radially from rest at infinity into a BH of mass
$M\gg m$ is $E=0.0104 m^2/M$~\cite{davis}. If $m$ is replaced by the
reduced mass $\mu$, this result agrees with the numerical result
$E\simeq 0.0013$ (see Fig.~14 and Sec.~IV of
Ref.~\cite{anninos-pert}). The slight discrepancy between
perturbative and numerical results can be quantitatively explained
by considering three fudge factors in the perturbative
analysis~\cite{anninos-pert}: $F_{r_0}$, which accounts for the
finite initial infall distance in the numerical simulations; $F_h$,
which accounts for tidal deformations heating up the BH horizon;
$F_{\rm abs}$, which accounts for the reabsorption of the
gravitational waves by the BH. In our case the process starts at
infinite separation, thus $F_{r_0}=1$, $F_h\simeq 0.86$, and $F_{\rm
abs}\simeq 0.99$.

Perturbation theory in the close-limit approximation and numerical simulations
are also consistent for four-dimensional ultra-relativistic collisions.
Numerical studies of boosted BHs with fixed initial separation show that the
energy emission saturates at $E\sim 0.01 M$ for very large initial BH momenta
(see, e.g., Fig.~2 of Ref.~\cite{boosted2}). Ref.~\cite{boosted} combines
Newtonian dynamics and numerical simulations of boosted BHs to conclude that
the maximum energy emission could actually be much lower than the above value,
$E\lesssim 0.0016 M$. Four-dimensional investigations also suggest that most of
the radiation is emitted in the ringdown phase. This is confirmed by comparing
perturbative calculations to post-Newtonian calculations \cite{simone}. The
bremsstrahlung radiation of a particle at distance larger than $4M$ (in
Schwarzschild coordinates) contributes only $\sim 3$ \% of the total energy
emitted.

In conclusion, numerical results in four dimensions indicate that
perturbation theory likely overestimates the emitted radiation. The
generalization of these results to higher dimensions is nontrivial,
and the values in Tables \ref{tab:efficiency2} and
{\ref{tab:efficiency3}} could underestimate the efficiency for
gravitational wave generation in high-energy collision of two equal
mass particles.

\section{\label{concl} Conclusions}
We reviewed and compared different approaches to computing the
gravitational energy released during back hole formation in
high-energy particle collisions. While in four dimensions there is
good agreement between all these methods, results differ
significantly in higher dimensions. The straightforward conclusion
is that new techniques and refinements in previous calculations are
needed to obtain a quantitative understanding of BH formation in
higher-dimensional spacetimes. This is particularly important for
high $D$, where different approaches do not yield consistent
results, and for non head-on collisions, where the trapped surface
method is expected to be less reliable.
\section*{Acknowledgements}
We warmly thank Slava Rychkov and Hirotaka Yoshino for a critical
reading of the manuscript and for many useful suggestions. V. C.
acknowledges financial support from Funda\c c\~ao para a Ci\^encia e
Tecnologia (FCT) - Portugal through grant SFRH/BPD/2004. This work
was supported in part by the National Science Foundation under grant
PHY 03-53180, and by a University of Mississippi FRP Grant.


\end{document}